\documentclass[preprint]{elsarticle}

\usepackage{amsfonts}
\usepackage{url}

\begin{document}

\title{Compartmental and cellular automaton $SEIRS$ epidemiology models for the COVID-19 pandemic with the effects of temporal immunity and vaccination}

\author{Jaroslav Ilnytskyi\corref{cor}}\ead{iln@icmp.lviv.ua}
\author{Taras Patsahan}
\address{Institute for Condensed Matter Physics of Nat. Acad. Sci. of Ukraine,\\ 1, Svientsitskii Str., Lviv, Ukraine}

\date{\today}

\begin{abstract}
We consider the $SEIRS$ epidemiology model with such features of the COVID-19 outbreak as: abundance of unidentified  infected individuals, limited time of immunity and a possibility of vaccination. Within a compartmental realization of this model, we found the disease-free and the endemic stationary states. They exist in their respective restricted regions of the le via linear stability analysis. The expression for the basic reproductive number is obtained as well. The positions and heights of a first peak for the fractions of infected individuals are obtained numerically and are fitted to simple algebraic forms, that depend on model rates. Computer simulations of a lattice-based realization for this model was performed by means of the cellular automaton algorithm. These allowed to study the effect of the quarantine measures explicitly, via changing the neighbourhood size. The attempt is made to match both quarantine and vaccination measures aimed on balanced solution for effective suppression of the pandemic.       
\end{abstract}

\begin{keyword}
epidemiology \sep cellular automata
%\PACS 83.80.Va \sep 61.41.+e \sep 61.20.Ja
\MSC: 92D30 \sep 37B15 \sep  92C60
\end{keyword}

\maketitle

%%%%%%%%%%%%%%%%%%%%%%%%%%%%%%%%
\section{\label{I}Introduction and the model}

At this stage of the spread of the COVID-19 epidemic, the qualitative impact of factors such as: the presence of a large number of asymptomatic cases, quarantine restrictions \cite{Kucharski2020, Rahimi2020}, loss of immunity to SARS-CoV-2 virus \cite{Lopez2020} due to its mutations \cite{Mallapaty2020, Hou2020, Goodman, Terry} and vaccination of the population, are generally well known from both numerous studies and real statistics data \cite{WHO_data}. However, many questions remain about the interaction of these factors and the choice of their optimal combination, both in terms of the most effective way to overcome the epidemic and with minimal economic losses. A wide range of theoretical and simulation approaches are used for this purpose \cite{Eren2020, Pongkitivanichkul2020, Gotz2020, Block2020,Blavatska2021}. In our previous work \cite{Ilnytskyi2021}, we considered the $SEIRS$ compartment model, which was investigated in detail in the limit of no immunity to the disease. This approximation makes sense for the case of the rapidly mutating virus and the lack of vaccination, which was a feature of the epidemic at the end of $2020$ and the beginning of $2021$. Since then, significant changes have taken place, in particular: extensive statistics have been collected on the characteristic duration of natural immunity, a number of vaccines have been developed with different principles of action, and others. These circumstances necessitate the inclusion of these factors in the consideration.

\begin{figure}[!ht]
\begin{center}
\includegraphics[clip,width=8cm,angle=0]{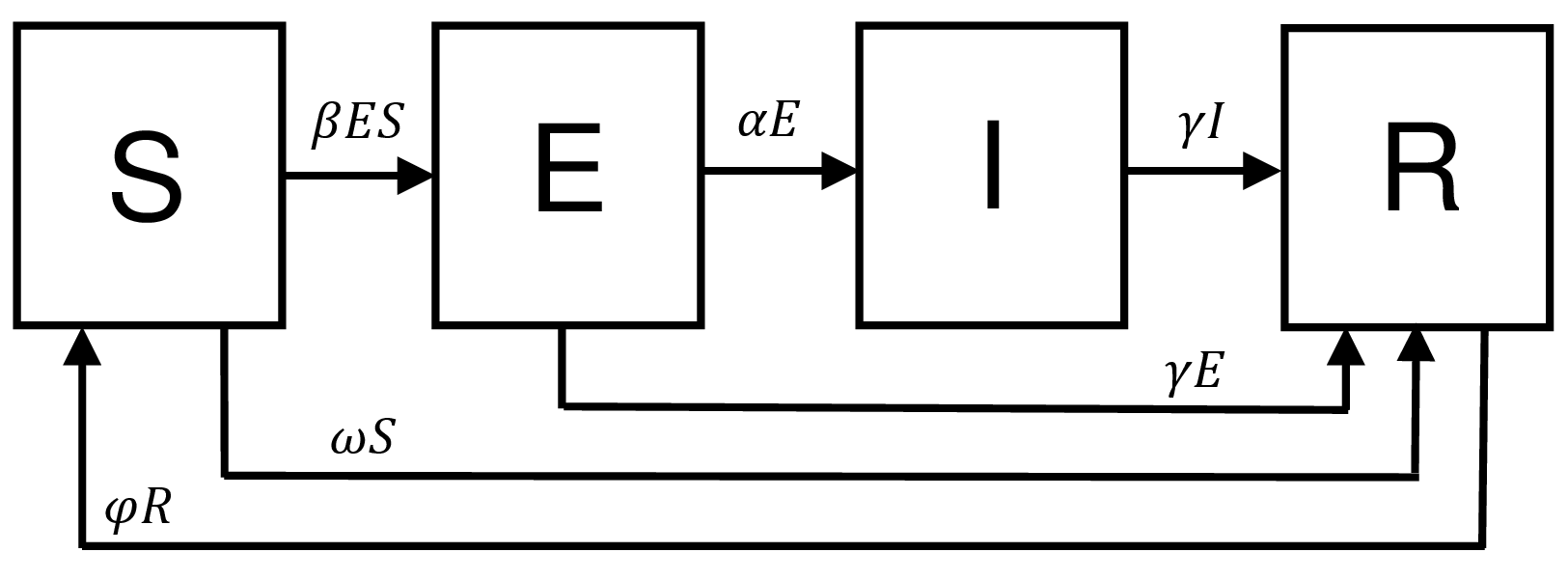}
\caption{\label{Model}The $SEIRS$ epidemiology model for the COVID-19 dissemination.}
\end{center}
\end{figure}

The $SEIRS$ compartmental model, suggested in this study and shown schematically in Fig.~\ref{Model}, contains four groups, marked via their respective fractions of susceptible $S$, unidentified infective $E$, identified isolated infective $I$, and recovered non-infective $R$. Group $E$ contains asymptomatic patients and those with mild or strong symptoms but untested and not isolated from the community. All these originate from the group $S$, turning into infective with the rate $\beta$. After their identification with the rate $\alpha$ using suitable tests, these are transfered further into the group $I$ of isolated (in home, or, in hospitals) infected individuals. Both unidentified and isolated identified individuals are supposed to lose their infectivity and to move into the recovered group $R$ with the same recovery rate $\gamma$. These are neither infective nor susceptible to the virus. This status lasts, however, not forever, and they are transfered into the susceptible group $S$ with the loss of immunity rate $\varphi$. The direct path from the $S$ to the $R$ group is possible via vaccination, with the vaccination rate $\omega$. To simplify the model, we assume negligible incubation period, instant isolation of identified infective individuals, and the birth and death rates being much smaller than the characteristic rates of the pandemic development.

Let us make a few more notes upon the relation of the coefficients $\beta$, $\alpha$, $\gamma$, $\varphi$ and $\omega$ to the real life. If the population size is $N$ individuals, then the absolute number of susceptible, unidentified infective, identified isolated infective and recovered non-infective individuals are $N_S=SN$, $N_E=EN$, $N_I=IN$ and $N_R=RN$, respectively. Assuming a time unit equal to a single day, the absolute number of individuals infected per day is $\beta N_S (N_E/N)$. This number can be interpreted in different ways, e.g. (i) each of $N_S$ susceptible individuals meets one of $N_E$ infective individuals per day with the probability of $N_E/N$ and, as the result, contracts a virus from him with the probability $\beta$, or (ii) by rewriting in a form $(\beta/k) N_S k(N_E/N)$, there were $k$ such contacts per day but the virus was contracted with the probability $\beta/k$ in each of them. Hence, two principal quarantine measures, such as: (a) intenisification of individual sanitary norms (masks, distance, frequent hand wash, etc.) aimed on reduction of the probability to contract the virus during the contact; and (b) restriction of contacts between individuals, can not be separated within the compartmental model formalism -- both are fused into a single coefficient $\beta$. The coefficient $\alpha$ reflects how widely the population is covered by appropriate medical testing and defines the number of newly identified infective individuals per day, $\alpha N_E$. Such type of data is typically published in an open press. Recovery rate $\gamma$, in the first approximation, can be taken as fixed for both $E$ and $I$ groups, reflecting an average loss of infectivity during 14 days, hence, $\gamma=1/14$. Likewise, the loss of immunity, which is currently debatable, can be fixed at about four months, yielding the coefficient $\varphi=1/120$. Finally, the coefficient $\omega$ reflects how widely the population is covered by vaccination and how effective the vaccine is, the absolute number of individuals vaccinated per day is $\omega N_S$. Here we neglect the vaccine activation period, assuming that it starts to act instantly. 

The corresponding set of differential equations has the following form:
\begin{eqnarray}
\dot{S} & = & -\beta ES - \omega S + \varphi R \label{dSdt}\\
\dot{E} & = & \beta ES - (\gamma + \alpha) E \label{dEdt}\\
\dot{I} & = & \alpha E - \gamma I \label{dIdt}\\
\dot{R} & = & \gamma(E+I) + \omega S - \varphi R \label{dRdt}\\
&& S+E+I+R  = 1\label{cond}
\end{eqnarray}

One can identify clearly the effect of the presence of isolated $I$ and recovered $R$ individuals in this model by introducing cumulative fractions of all uninfected individuals, $S'=S+R$, and all infected ones, $E'=E+I$, where $S'+E'=1$ holds. The equations set (\ref{dSdt})-(\ref{dRdt}) can be rewritten as a single equation for $E'$
\begin{equation}
\dot{E'} = \beta(E'-I)(S'-R) - \gamma E', \label{dEpdt}
\end{equation}
which has the same form as the one for the $SIS$ epidemiology model \cite{Ilnytskyi2016}, when $S$ and $I$ are substituted via $S'$ and $E'$, respectively, but with the reduced number of transmission acts. The latter are given by the product $(E'-I)(S'-R)$, where both infective and susceptible parties are reduced by $I$ and $R$, respectively. The equation (\ref{dEpdt}) is, obviously, not self-sufficient, as one needs to complement it by the equations providing the time evolution for the $I$ and $R$ fractions. To do so, one may attempt either to rewrite the equations set (\ref{dSdt})-(\ref{dRdt}) in terms of four variables $S'$, $E'$, $I$ and $R$, or to use certain approximations to express $E$ and $R$ fractions via $S'$ and $I'$ ones, similarly as suggested in Ref.~\cite{Ilnytskyi2021} for the $SEIRS$ model without the immunity and vaccination considered there.

The purpose of this study is to examine the stationary states and the time evolution of the $SEIRS$ model as defined in Fig.~\ref{Model} with the emphasis on the quarantine measures, the role of immunity loss and vaccination on pandemic dynamics. The study is of a general type with no direct link to particular country/region or statistical data of any sort. We, therefore, focus on features and tendencies as predicted by this model and not on practical recommendations that can be used straightaway. Section \ref{II} contains analysis of the stationary states (fixed points) for the model and analysis of their stability; in section \ref{III} we discuss early-time spread and the decay dynamics of the disease dissemination by combining numerical and approximate analytic tools; in section \ref{IV} we consider the effects for quarantine measures and their relaxation on the dynamics of the COVID-19 pandemic, especially on the height of the second wave of the disease, section \ref{V} contains conclusions.

%%%%%%%%%%%%%%%%%%%%%%%%%%%%%%%%
\section{Fixed points and their stability}\label {II}

The stationary state (fixed points) for the $SEIRS$ model is given as the solution of the equations set:
\begin{eqnarray}
&& -\beta ES - \omega S + \varphi R = 0 \label{Sfp}\\
&& \beta ES - (\gamma + \alpha) E = 0 \label{Efp}\\
&& \alpha E - \gamma I = 0 \label{Ifp}\\
&& \gamma(E+I) + \omega S - \varphi R = 0 \label{Rfp}\\
&& S + E + I + R = 1 \label{cond_fp}
\end{eqnarray}
We will restrict our analysis to the case when both the loss of infectivity, $\gamma$, and the loss of immunity, $\varphi$, rates are constant and non-zero 
\begin{equation}\label{gp_nonzero}
\gamma=\mathrm{const.}>0,\hspace{2em} \varphi=\mathrm{const.}>0.
\end{equation}

\begin{itemize}
\item Disease-free (DF) fixed point
\end{itemize}

We will denote hereafter all the fractions in the DF fixed point by the $\dagger$ superscript. The definition of the DF fixed point requires that $E^{\dagger}=I^{\dagger}=0$. It is easy to see from the Eq.~(\ref{Ifp}) that, if condition (\ref{gp_nonzero}) holds, then $I^{\dagger}=(\alpha/\gamma) E^{\dagger}$ and both these fractions turns into zero simultaneously. The set of equations is reduced to
\begin{eqnarray}
&& \omega S^{\dagger} - \varphi R^{\dagger} = 0 \label{dfeq1}\\
&& S^{\dagger} + R^{\dagger} = 1 \label{dfeq2}
\end{eqnarray}
yielding the solution:
\begin{equation}\label{DF_sol}
S^{\dagger}=\frac{\varphi}{\varphi+\omega},~~E^{\dagger}=I^{\dagger}=0,~~R^{\dagger}=\frac{\omega}{\varphi+\omega} 
\end{equation}

\begin{itemize}
\item Endemic (EN) fixed point
\end{itemize}

The fractions in the EN fixed point will be denoted by the $^*$ superscript. The endemic fixed point is defined as such, that both $E^*>0$ and $I^*>0$. They are related via Eq.~(\ref{Ifp}) resulting in $I^*=(\alpha/\gamma) E^*$, with both $\gamma>0$ and $\alpha>0$, hence, a single condition $E^*>0$ is sufficient. In this case, both sides of Eq.~(\ref{Efp}) can be divided by $E^*$ resulting in a straightaway solution for $S^*=(\gamma+\alpha)/\beta$ for $S^*$. Now, the remaining equations in the set (\ref{Sfp})-(\ref{cond_fp}) are
\begin{eqnarray}
&& (\gamma+\alpha)E^* - \varphi R^* = - \omega S^* \label{eneq1}\\
&& (\gamma+\alpha)E^* + \gamma R^* = \gamma (1-S^*) \label{dfeq2}
\end{eqnarray}
The complete solution for the EN fixed point can be written in the following form
\begin{eqnarray}
S^* & = & \frac{\gamma+\alpha}{\beta} \label{EN_sol_S}\\ 
E^* & = & \frac{\gamma\varphi}{(\gamma+\alpha)(\gamma+\varphi)}\left[1-S^*/S^{\dagger}\right] \label{EN_sol_E}\\
I^* & = & \frac{\alpha\varphi}{(\gamma+\alpha)(\gamma+\varphi)}\left[1-S^*/S^{\dagger}\right] \label{EN_sol_I}\\
R^* & = & \frac{\gamma}{\gamma+\varphi}\left[1-(1-\frac{\omega}{\gamma})S^*\right] \label{EN_sol_R}
\end{eqnarray}
Again, this solution exists only at $E^*>0$, which, according to Eq.~(\ref{EN_sol_E}) yields $S^*<S^\dagger$ for the $S^*$ fraction. Following Eq.~(\ref{EN_sol_S}), one can see that, if the expression for $S^*=\frac{\gamma+\alpha}{\beta}$ became equal or greater than $S^\dagger$, the crossover to the DF fixed point occurs. Using the expression for $S^\dagger$ (\ref{DF_sol}), this condition for the DF fixed point can be written as
\begin{equation}\label{R0}
R_0=\frac{\beta\varphi}{(\gamma+\alpha)(\varphi+\omega)} \leq 1,
\end{equation}
where $R_0$ has a meaning of the basic reproductive number. Let us note, that in the limit case of $\alpha=\omega=0$ (no identification and no vaccination), the expression $R_0=\beta/\gamma$ for the $SIS$ model is retrieved (in this case the $E$ fraction of the current model serves as the $I$ fraction in the $SIS$ model). Taking into account the assumption (\ref{gp_nonzero}), the expression (\ref{R0}) indicates the way of bringing the basic reproductive number $R_0$ down by means of both decrease of the transmission rate $\beta$ and by the increase of the identification $\alpha$ and vaccination $\omega$ rates. One can rewrite the condition $R_0 \leq 1$ for the DF state in any of three following forms
\begin{equation}\label{bc_ac_oc}
\beta\leq\beta_c=\frac{(\gamma+\alpha)(\varphi+\omega)}{\varphi},~~\alpha\geq\alpha_c=\frac{\beta\varphi}{\varphi+\omega}-\gamma,~~\omega\geq\omega_c=\frac{\beta\varphi}{\gamma+\alpha}-\varphi
\end{equation}
These simple relations have much practical use. For instance, at any fixed identification $\alpha$ and vaccination $\omega$ rates, there is a maximum allowed transmission rate $\beta_c$ at which the pandemic can be brought down. If current transmission rate $\beta>\beta_c$, more strict quarantine and hygiene measures should be introduced. Similarly, at current transmission $\beta$ and vaccination $\omega$ rates, there is a minimal identification rate $\alpha_c$ at which the pandemic can be defeated. If current identification is made with a slower rate, then it should be increased. The same can be said about the critical vaccination rate $\omega_c$.

Another useful consequence of obtaining Eq.~(\ref{R0}) is the possibility to obtain the expression for its full differential
\begin{equation}\label{dR0}
\frac{1}{R_0}{dR_0}=\frac{1}{\beta}{d\beta} - \frac{1}{\gamma+\alpha}{d\alpha} - \frac{1}{\varphi+\omega}{d\omega}   
\end{equation}
It provides the respective weights, with which the infinitesimal changes of all variable model parameters, $d\beta$, $d\alpha$ and $d\omega$, affect the resulting infinitesimal change $dR_0/R_0$ of the basic reproductive number. If the financial burden, associated with all of $d\beta$, $d\alpha$ and $d\omega$ can be estimated, then the optimal strategy of the most economic way to bring the pandemic down can be evaluated.

Let us concentrate now on linear stability analysis for both fixed points. To reduce the number of parameters, we eliminate the $R$ fraction by using the expression (\ref{cond}). Then, the equations set (\ref{dSdt})-(\ref{cond}) can be rewritten as
\begin{eqnarray}
\dot{S} & = & -\beta ES - \omega S + \varphi (1-S-E-I)\label{dSdt_st}\\
\dot{E} & = & \beta ES - (\gamma + \alpha) E \label{dEdt_st}\\
\dot{I} & = & \alpha E - \gamma I \label{dIdt_st}
\end{eqnarray}
The eigenvalues $\lambda$ of the Jacobian matrix $\mathbf{J}$ are given by the equation
\begin{equation}
\det \mathbf{J}=\left|
\begin{array}{ccc}
-\beta E-(\varphi+\omega)-\lambda & -\beta S-\varphi& -\varphi\\
\beta E & \beta S-(\gamma+\alpha)-\lambda & 0\\
0 & \alpha & -\gamma-\lambda
\end{array}
\right|=0
\end{equation}
The determinant can be reduced using the first column leading to the equation of the form
\begin{equation}\label{lambda_eqs}
(\lambda+\gamma)(\lambda+\beta E+\varphi+\omega)(\lambda-\beta S+\gamma+\alpha) + \beta E \left[(\lambda+\gamma)(\beta S + \varphi) + \alpha\varphi\right] = 0
\end{equation}

For the DF fixed point, we substitute $\{S,E\}$ by $\{S^\dagger,E^\dagger\}$, the expressions for the latter are given by Eq.~(\ref{DF_sol}). This simplifies the equation to
\begin{equation}\label{lambda_eqs_DF}
(\lambda+\gamma)(\lambda+\varphi+\omega)(\lambda-\beta S^\dagger+\gamma+\alpha) = 0
\end{equation}
and provide the roots: $\lambda^\dagger_1=-\gamma$, $\lambda^\dagger_2=-(\varphi+\omega)$, and $\lambda^\dagger_3=\beta S^\dagger - (\gamma+\alpha)=-(\gamma+\alpha)(1-R_0)$. Because $\gamma > 0$ and $\varphi+\omega > 0$, then both $\lambda^\dagger_1$ and $\lambda^\dagger_2$ are always negative. One has $R_0\leq1$ in the DF fixed point, hence, $\lambda^\dagger_3\leq 0$ there. Therefore, the linear stability analysis indicates the DF fixed point to be a stable one at $R_0<0$ and cannot provide the answer of its stability at $R_0=1$. 

For the EN fixed point, we substitute $\{S,E\}$ by $\{S^*,E^*\}$, the expressions for the latter are given by Eqs.~(\ref{EN_sol_S}) and (\ref{EN_sol_E}). From there one sees that $\beta S^*=\gamma+\alpha$, and the Eq.~(\ref{lambda_eqs}) takes the following form
\begin{equation}\label{lambda_eqs_EN}
\lambda(\lambda+\gamma)(\lambda+\beta E^*+\varphi+\omega) + \beta E^* \left[(\lambda+\gamma)(\gamma+\alpha+\varphi) + \alpha\varphi\right] = 0
\end{equation}
It can be rewritten as
\begin{displaymath}
\lambda^3 + \left[\gamma+\varphi+\omega+\beta E^*\right]\lambda^2 + \left[\gamma(\varphi+\omega)+(\gamma+\alpha+\gamma+\varphi)\beta E^*\right]\lambda
\end{displaymath}
\begin{equation}\label{lambda_eqs_EN_2}
 ~~~~~~~~~~~~~~~~~~~~~~~~~~~~~~~+(\gamma+\alpha)(\gamma+\varphi)\beta E^* = 0
\end{equation}
It will be examined graphically, in exactly the same way as for the case of the $SEIRS$ model with no immunity \cite{Ilnytskyi2021}, hence we omit handbook details \cite{korn2013mathematical} here. There are three variable parameters, $\alpha$, $\beta$ and $\varphi$, therefore we will consider the discriminant $Q$ of the cubic equation (\ref{lambda_eqs_EN_2}) and its three roots, $\lambda_1$, $\lambda_2$ and $\lambda_3$, as the functions of $\alpha$ and $\beta$ at fixed vaccination rate $\omega$.

\begin{figure}[!ht]
\begin{center}
\includegraphics[clip,width=12cm,angle=0]{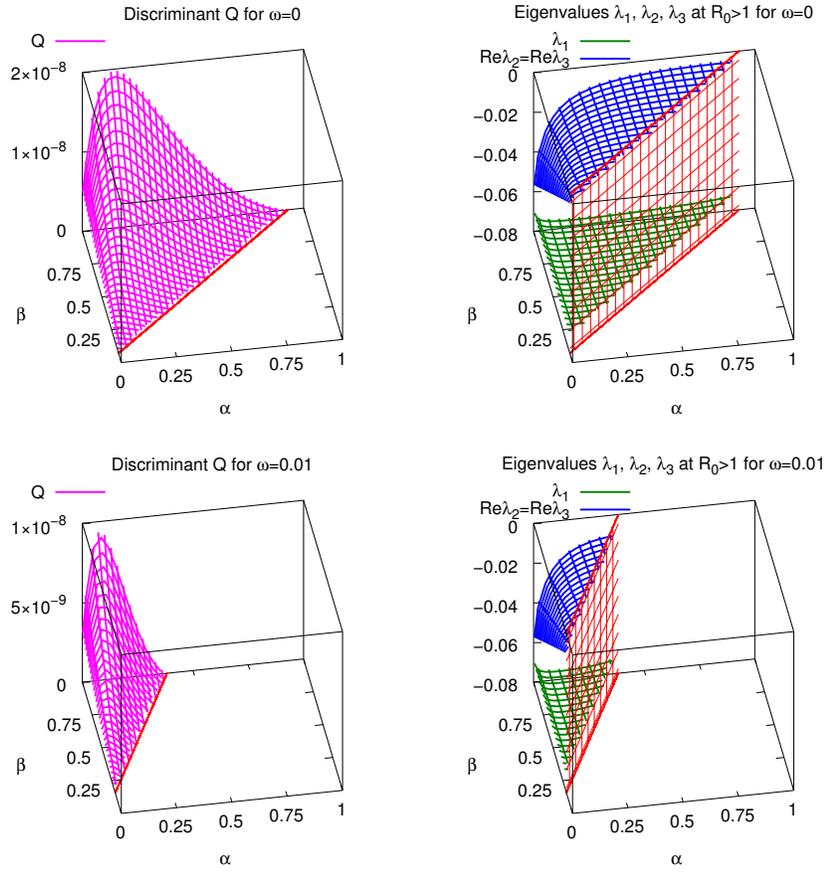}
\caption{\label{Q_lambda}Top row: the discriminant $Q$ of the cubic equation (\ref{lambda_eqs_EN_2}) and the real parts of its roots $\lambda_i$ for the no vaccination case of $\omega=0$. Bottom row: the same for the vaccination rate of $\omega=0.01$.}
\end{center}
\end{figure}
The no vaccination case is shown in the top row of Fig.~\ref{Q_lambda} and it will be considered first. All quantities of interest, $Q$ and $\lambda_i$, are shown only within the region $(\alpha,\beta)$ characterized by $R_0>1$, where the EN fixed point (\ref{EN_sol_S})-(\ref{EN_sol_R}) is a valid solution. The crossover to the DF fixed point occurs along the line given by $R_0=1$, shown in red in the surface plot for $Q$. For the sake of clarity, it is translated along the $Z$-axis into the red dashed wall in the surface plots for $\lambda_i$. The surface representing $Q$, indicates positive values for the latter in the whole $R_0>1$ region, hence, the Eq.~(\ref{lambda_eqs_EN_2}) has one real, $\lambda_1$, and two complex, $\lambda_{2}$ and $\lambda_{3}$, roots. As follows from the surface plots for $\lambda_1$ and for the real parts of $\lambda_{2}$ and $\lambda_{3}$, all three are negative in the whole $R_0>1$ region. This confirms stability of the EN fixed point within this region, examined in a graphical way.

The case of the vaccination rate $\omega=0.01$ is shown in the second row of Fig.~\ref{Q_lambda}. One observes rotation of the $R_0$ line here and reduction of the area for the $R_0>1$ region, as compared to the $\omega=0$ case. The surface plot for $Q$ keeps its general shape, but is more ``jammed'' from the crossover line towards the $(\alpha=1,\beta=1)$ point. The values of $Q$ keep remaining positive in the whole $R_0>1$ region. In contrast to this, the surfaces for $\lambda_1$ and for the real parts of $\lambda_{2}$ and $\lambda_{3}$, appear to be the same as in the case of $\omega=0$, but are cut now at new position of the crossover line $R_0=1$. All three are negative within the $R_0>1$ region. The same holds true upon the further increase of the vaccination rate $\omega$ (not shown), until the $R_0>1$ region moves out of the square defined by the $\alpha\in[0:1]$ and $\beta\in[0:1]$ boundaries. As the result, we conclude that the EN fixed point is stable within the $R_0>1$ region at all vaccination rates $\omega$.
    
%%%%%%%%%%%%%%%%%%%%%%%%%%%%%%%%
\section{Numeric solution and approximate expressions for the first peak position and height}\label {III}

Here we employ numerical integration of Eqs.~(\ref{dSdt})-(\ref{dRdt}), performed via the second-order integrator
\begin{equation}
X(t+\Delta t)=X(t) + \dot{X}(t)\Delta t + \frac{1}{2}\ddot{X}(t)\Delta t^2\label{integr}
\end{equation}
for each fraction $X=\{S,E,I,R\}$, applied iteratively with the time step $\Delta t$ equal to one day. At time instance $t=0$, the system is characterized by the fraction of unidentified infected individuals $E(0)=E_0$, assumed to be brought from outside, and the other fractions are: $S(0)=1-E_0$ and $I(0)=R(0)=0$. Different values of $E_0$ are examined. The equations are coupled, as far as both the first derivatives $\dot{X}$, given by Eqs.~(\ref{dSdt})-(\ref{dRdt}), and the second derivatives
%
%    d2S = -beta*dE*S - beta*E*dS - omega*dS + phi*dR
%    d2E = beta*dE*S + beta*E*dS  - (gamma+alpha)*dE
%    d2I = alpha*dE - gamma*dI
%    d2R = gamma*(dE+dI) + omega*dS - phi*dR
\begin{eqnarray}
\ddot{S} & = & -\beta(S\dot{E} + E\dot{S}) - \omega \dot{S} + \varphi \dot{R} \label{ddSdt}\\
\ddot{E} & = & \beta(S\dot{E}+ E\dot{S}) - (\gamma + \alpha) \dot{E} \label{ddEdt}\\
\ddot{I} & = & \alpha \dot{E} - \gamma \dot{I} \label{ddIdt}\\
\ddot{R} & = & \gamma (\dot{E}+\dot{I}) + \omega \dot{S} - \varphi \dot{R} \label{ddRdt}
\end{eqnarray}
at time instance $t$ depends on all variables $S$, $E$, $I$ and $R$ at the same instance $t$. The source of infection is the $E$ fraction, because in the current model, depicted in Fig.~\ref{Model}, we consider an ideal case of instant isolation of identified infective individuals into group $I$. On the other hand, part of the latter are isolated in the hospitals, therefore $I$ reflects the load put on the health system. Therefore, we concentrate on the time evolutions, $E(t)$ and $I(t)$ of these two fractions. These are examined at various initial values of $E_0$ and at various $\beta$, $\alpha$ and $\omega$ rates.

%%%%%
\subsection{The no vaccination case, $\omega=0$}

\begin{figure}[!ht]
\begin{center}
\includegraphics[clip,width=12cm,angle=0]{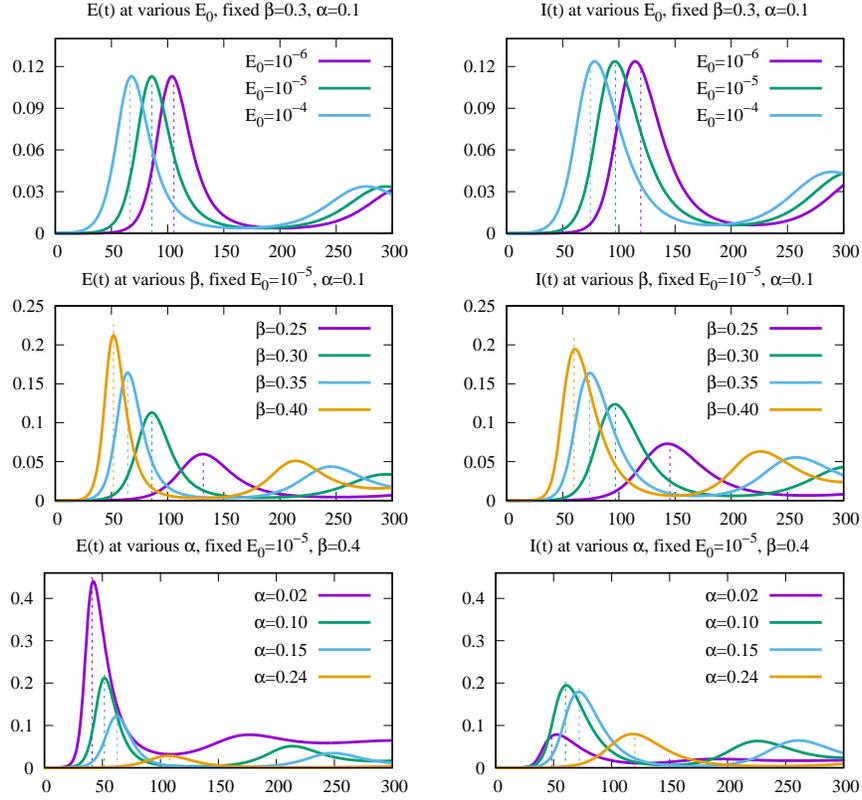}
\caption{\label{novacc_EI_evol}Time evolutions of the unidentified infected, $E(t)$, and isolated infective, $I(t)$, fractions depending on variation of the model parameters, as provided by a numeric solution. The no vaccination case $\omega=0$ is shown. Top row: the effect of variation of the initial value $E_0$ at fixed $\beta$ and $\alpha$; middle row: the same for the contact rate $\beta$ at fixed $E_0$ and $\alpha$; bottom row: the same for the identification rate $\alpha$ at fixed $E_0$ and $\beta$. Dashed vertical lines show approximate positions and heights of the first peak and are the results of approximate analytic expressions, see explanation further in the text.}
\end{center}
\end{figure}
We will discuss the no vaccination case ($\omega=0$) first. In contrary to the $SEIRS$ model with no immunity \cite{Ilnytskyi2021}, we observe oscillatory behaviour for $E(t)$ and $I(t)$ for most combinations of $\beta$ and $\alpha$ rates being considered. Both plots at the top row of Fig.~\ref{novacc_EI_evol} indicate that, at fixed $\beta$ and $\alpha$, the decrease of $E_0$ does not affect the first peak heights, $E_{\mathrm{max}}$ and $I_{\mathrm{max}}$, for both fractions, but shifts their respective positions, $t_{\mathrm{max},E}$ and $t_{\mathrm{max},I}$, towards the later times. The amount of a shift is found to be proportional to $-\log E_0$. The plots displayed in a middle row shows that the decrease of a contact rate $\beta$ decreases both $E_{\mathrm{max}}$ and $I_{\mathrm{max}}$ and shifts $t_{\mathrm{max},E}$ and $t_{\mathrm{max},I}$. The same effect is achieved by the increase of the $\alpha$ rate, as shown in the bottom frame of Fig.~\ref{novacc_EI_evol}.

\begin{figure}[!ht]
\begin{center}
\includegraphics[clip,width=10cm,angle=0]{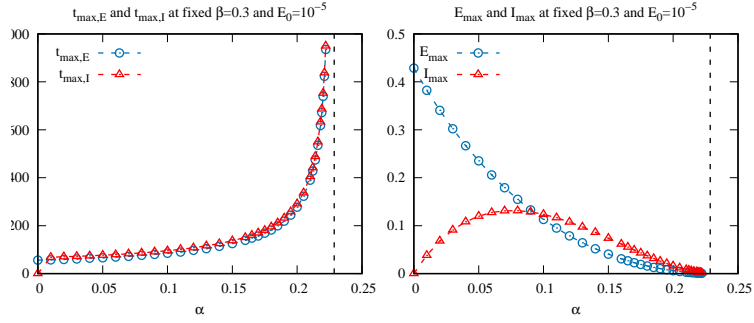}
\caption{\label{tmax_EImax_ex}Example for the $t_{\mathrm{max},E}$, $t_{\mathrm{max},I}$, $E_{\mathrm{max}}$ and $I_{\mathrm{max}}$ as the functions of the identification rate $\alpha$ at fixed transmission rate $\beta=0.3$ and initial condition $E_0=10^{-5}$. Vertical dashed lines provide the position of the critical identification rate $\alpha_c$, given by Eq.~(\ref{bc_ac_oc}).}
\end{center}
\end{figure}
Both the first peak heights, $E_{\mathrm{max}}$ and $I_{\mathrm{max}}$, and their respective positions, $t_{\mathrm{max},E}$ and $t_{\mathrm{max},I}$, are of great practical interest for predicting of the pandemic development. It is, however, impractical to solve Eqs.~(\ref{dSdt})-(\ref{dRdt}) numerically at each required parameters set for this purpose, and one would improve the predictive practicality of the modelling approach by suggesting simple approximate expressions instead. To this end we examined a general shape for all properties of interest, $t_{\mathrm{max},E}$, $t_{\mathrm{max},I}$, $E_{\mathrm{max}}$ and $I_{\mathrm{max}}$, as the functions of the identification rate $\alpha$ at various fixed transmission rate $\beta$ and initial conditions given by $E_0$. We found that both peak positions, $t_{\mathrm{max},E}$ and $t_{\mathrm{max},I}$, diverge as $\alpha$ approaches $\alpha_c$, where the latter is defined in Eq.~(\ref{bc_ac_oc}), see left frame in Fig.~\ref{tmax_EImax_ex}. Both peak heights, $E_{\mathrm{max}}$ and $I_{\mathrm{max}}$, decay to zero as $\alpha$ approaches $\alpha_c$, see right frame in the same plot. These observations, alongside with the one from Fig.~\ref{novacc_EI_evol}, that both $t_{\mathrm{max},E}$ and $t_{\mathrm{max},I}$ are proportional to $-\log E_0$ and $E_{\mathrm{max}}$ and $I_{\mathrm{max}}$ are independent on $E_0$, led to suggesting scaling expressions of the following form
\begin{eqnarray}
t_{\mathrm{max},E}(\alpha,\beta,E_0) &=& -\log(E_0)A(\beta)(1-\frac{\alpha}{\alpha_c(\beta)})^{-v(\beta,E_0)},\label{tmaxE}\\
t_{\mathrm{max},I}(\alpha,\beta,E_0) &=& -\log(E_0)B(\beta)(1-\frac{\alpha}{\alpha_c(\beta)})^{-w(\beta,E_0)},\label{tmaxI}\\
E_{\mathrm{max}}(\alpha,\beta) &=& C(\beta)(1-\frac{\alpha}{\alpha_c(\beta)})^{p(\beta)},\label{Emax}\\
I_{\mathrm{max}}(\alpha,\beta) &=& D(\beta)\frac{\alpha}{\alpha_c(\beta)}(1-\frac{\alpha}{\alpha_c(\beta)})^{q(\beta)},\label{Imax}\\
&&\mathrm{where}~~ \alpha_c(\beta) = \beta - \gamma,\label{alphac}
\end{eqnarray}
Functional forms for all unknown coefficients, $A(\beta)$, $B(\beta)$, $C(\beta)$ and $D(\beta)$, and related exponents, $v(\beta,E_0)$, $w(\beta,E_0)$, $p(\beta)$ and $q(\beta)$, are obtained by fitting the data obtained by the numeric solution given by expression (\ref{integr}). Let us remark, that, contrary to the theory of phase transitions \cite{PhTr}, where one seeks the universal critical exponents in the vicinity of critical point (in our case, when $\alpha\to\alpha_c$), we opted here for so-called effective critical exponents, that are able to approximated the properties of interest in a wider range of the $\alpha$ rate. Therefore, the exponents $v$, $w$, $p$ and $q$ are the functions of either $\beta$ or both $\beta$ and $E_0$.

\begin{figure}[!ht]
\begin{center}
\includegraphics[clip,width=12cm,angle=0]{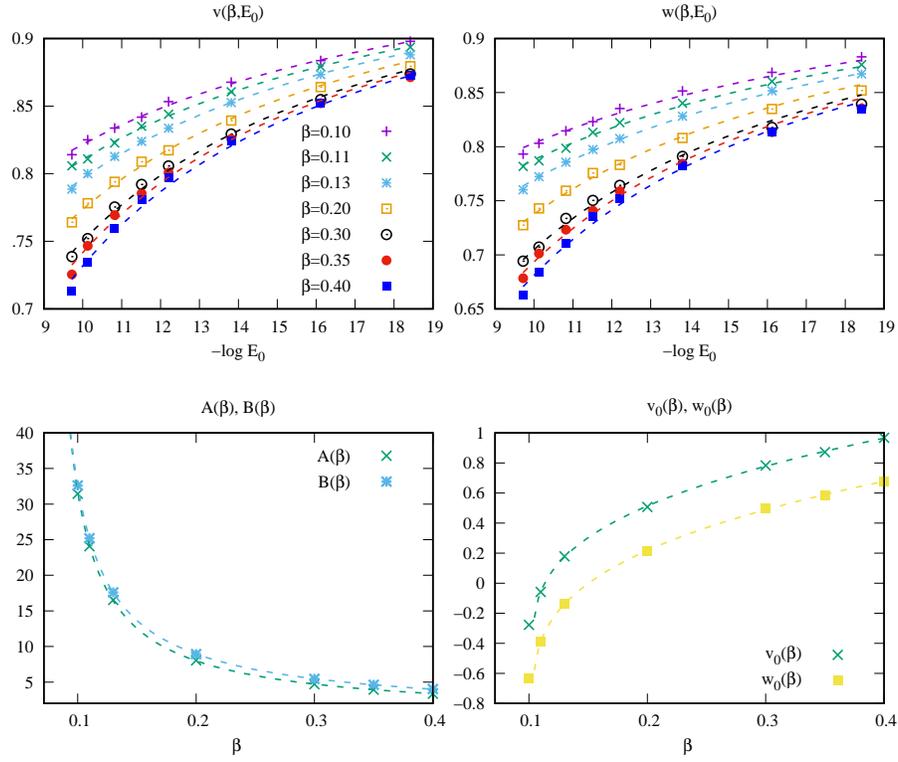}
\caption{\label{tmax_coeff}Coefficients $A{\beta}$, $B{\beta}$, and the exponents, $v{\beta,E_0}$, $w{\beta,E_0}$, for the scaling expressions (\ref{tmaxE}) and (\ref{tmaxI}).}
\end{center}
\end{figure}
The peak heights coefficients and exponents, as the functions of their respective arguments, are shown in Fig.~\ref{tmax_coeff}. They are fitted by the following expressions
\begin{eqnarray}
A(\beta)&=&1.22/(\beta-\gamma)^{0.92}\label{A},\\
v(\beta,E_0) &=& \frac{2}{\pi}\arctan[-v_0(\beta) + 0.32(-\log E_0)],\label{v}\\
v_0(\beta)&=&-0.40+2.11(\beta-0.104)^{0.36}\label{v0},\\
B(\beta)&=&1.54/(\beta-\gamma)^{0.86}\label{B},\\
w(\beta,E_0) &=& \frac{2}{\pi}\arctan[-w_0(\beta) + 0.25(-\log E_0)],\label{w}\\
w_0(\beta)&=&-0.81+2.22(\beta-0.104)^{0.33}\label{w0}.
\end{eqnarray}
\begin{figure}[!ht]
\begin{center}
\includegraphics[clip,width=9cm,angle=0]{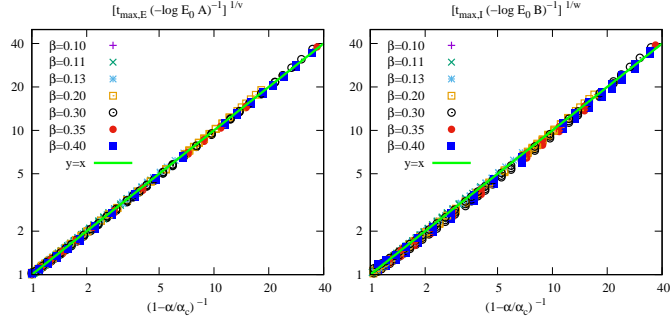}
\caption{\label{tmax_scl}Accuracy check for the scaling expressions (\ref{tmaxE}) and (\ref{tmaxI}). A wide range of transmission rates, $\beta\in[0.1:0.4]$, is displayed, where for each $\beta$ three different initial conditions, $E_0=10^{-7}$, $10^{-6}$ and $10^{-5}$ are studied and displayed in the plot.}
\end{center}
\end{figure}
As an accuracy check, we display the scaling plots for the combinations $\left[t_{\mathrm{max},E}/(-\log(E_0)A)\right]^{1/v}$ and $\left[t_{\mathrm{max},I}/(-\log(E_0)B\right]^{1/w}$ vs $(1-\alpha/\alpha_c)^{-1}$, obtained at arrange of values of $\beta$ and $E_0$, see Fig.~\ref{tmax_scl}. The data, obtained by means of numeric integration, is found to follow the expected $y(x)=x$ dependence very closely.

\begin{figure}[!ht]
\begin{center}
\includegraphics[clip,width=9cm,angle=0]{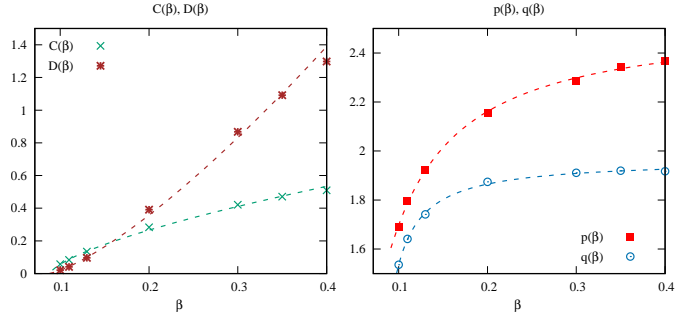}
\caption{\label{EImax_coeff}Coefficients $C{\beta}$, $D{\beta}$, and the exponents, $p{\beta,E_0}$, $q{\beta,E_0}$, for the scaling expressions (\ref{Emax}) and (\ref{Imax}).}
\end{center}
\end{figure}
\begin{figure}[!ht]
\begin{center}
\includegraphics[clip,width=9cm,angle=0]{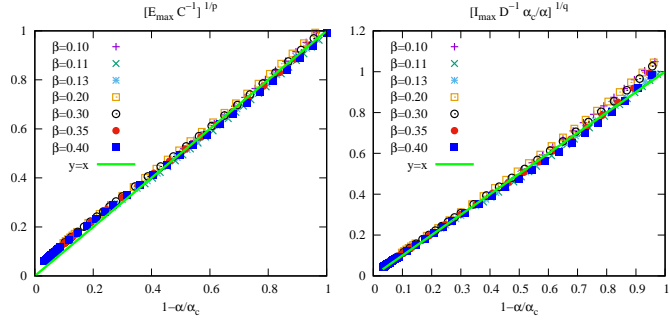}
\caption{\label{EImax_scl}The same as in Fig.~\ref{tmax_scl} but for the scaling expressions (\ref{Emax}) and (\ref{Imax}).}
\end{center}
\end{figure}

In a similar way, the peak positions coefficients and exponents, as the functions of their respective arguments, are shown in Fig.~\ref{EImax_coeff}. They are fitted by the following expressions
\begin{eqnarray}
C(\beta)&=&1.18/(\beta-0.089)^{0.68}\label{C}\\
p(\beta) &=& 1.62\arctan(-0.70+24.46\beta),\label{p}\\
D(\beta)&=&6.53/(\beta-0.089)^{1.33}\label{D}\\
q(\beta) &=& 1.25\arctan[-6.89+96.82\beta],\label{q}
\end{eqnarray}
And again, as an accuracy check, we display the scaling plots for the combinations $[E_{\mathrm{max}}/C]^{p}$ and $[I_{\mathrm{max}}\alpha_c/\alpha/D]^{q}$ vs $1-\alpha/\alpha_c$ in Fig.~\ref{EImax_scl} and found these to follow the $y(x)=x$ dependence reasonably well, too.

%

%%%%%
\subsection{The vaccination case, $\omega>0$}

\begin{figure}[!ht]
\begin{center}
\includegraphics[clip,width=12.3cm,angle=0]{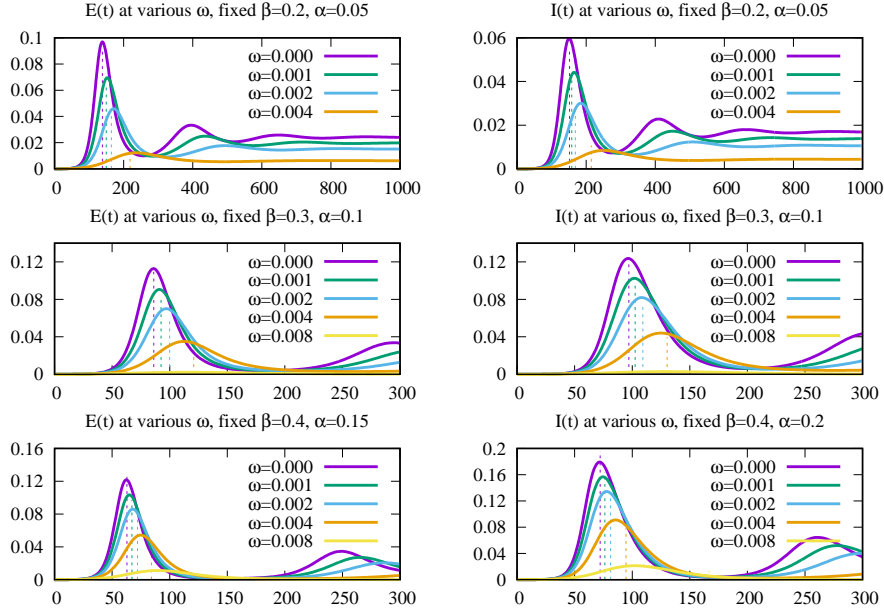}
\caption{\label{vacc_EI_evol}The effect of vaccination rate $\omega$ on time evolution of the fractions $E(t)$ and $I(t)$ at various values for the $\alpha$ and $\beta$ rates. Dashed vertical lines show approximate positions and heights of the first peak and are the results of approximate analytic expressions, see explanation in the text.}
\end{center}
\end{figure}
Similar analysis was performed for the case, when population was vaccinated during the epidemics spread with the rate $\omega$. Vaccination is evidenced via shifts for the first peak positions for both $E(t)$ and $I(t)$ fractions; and via essential decrease of their heights. These effects are illustrated in Fig.~\ref{vacc_EI_evol} for the numeric solutions for both fractions, where the approximate estimates for the peak positions and heights are also shown via vertical dashed lines. The approximate expressions read
\begin{eqnarray}
t_{\mathrm{max},E}(\alpha,\beta,E_0,\omega) &=& -\log(E_0)A(\beta)\tau_E(\beta,\omega)(1-\frac{\alpha}{\alpha^E_c(\beta,\omega)})^{-v(\beta,E_0)},\label{tmaxE}\\
t_{\mathrm{max},I}(\alpha,\beta,E_0,\omega) &=& -\log(E_0)B(\beta)\tau_I(\beta,\omega)(1-\frac{\alpha}{\alpha^I_c(\beta,\omega)})^{-w(\beta,E_0)},\label{tmaxI}\\
E_{\mathrm{max}}(\alpha,\beta,\omega) &=& C(\beta)\mu_E(\beta,\omega)(1-\frac{\alpha}{\alpha^E_c(\beta,\omega)})^{p(\beta)},\label{Emax}\\
I_{\mathrm{max}}(\alpha,\beta,\omega) &=& D(\beta)\mu_I(\beta,\omega)\frac{\alpha}{\alpha^I_c(\beta,\omega)}(1-\frac{\alpha}{\alpha^I_c(\beta,\omega)})^{q(\beta)},\label{Imax}\\
&&\mathrm{where}~~ \alpha^E_c(\beta,\omega) = (\frac{\beta\varphi}{\omega+\varphi} - \gamma)\exp(45\omega)\label{alphaEc}\\
&&\mathrm{and}~~ \alpha^I_c(\beta,\omega) = (\frac{\beta\varphi}{\omega+\varphi} - \gamma)\exp(35\omega).\label{alphaIc}
\end{eqnarray}
The expressions for $A(\beta)$, $v(\beta,E_0)$, $B(\beta)$, $w(\beta,E_0)$, $C(\beta)$, $p(\beta)$, $D(\beta)$ and $q(\beta)$ have been obtained above, see Eqs.~(\ref{A})-(\ref{q}). Critical value for $\alpha$ depends on $\omega$ now and the expressions for $\alpha^E_c(\beta,\omega)$ and $\alpha^I_c(\beta,\omega)$ differ by the exponent argument prefactor only. The no vaccination limit case, $\alpha^E_c(\beta,0)=\alpha^I_c(\beta,0)=\alpha_c(\beta)$, where $\alpha_c(\beta)$ is given by Eq.~(\ref{alphac}), holds. On a top of the dependence of the critical value on $\omega$, both peak positions and heights in Eqs.~(\ref{tmaxE})-(\ref{Imax}) acquires additional, $\omega$-dependent, multiplier,
\begin{eqnarray}
\tau_E(\beta,\omega) &=& \exp(-5\omega(\beta-\gamma)),\label{tauE}\\ 
\tau_I(\beta,\omega) &=& \exp(-\frac{5\omega}{\beta-\gamma}),\label{tauI}\\ 
\mu_E(\beta,\omega) &=& \exp(-\frac{19\omega}{\beta-\gamma}),\label{muE}\\ 
\mu_I(\beta,\omega) &=& \exp(-\frac{32\omega}{\beta-\gamma}),\label{muI} 
\end{eqnarray}
respectively. All these multipliers disappear at the no vaccination limit $\omega=0$.

The approximate expressions (\ref{tmaxE})-(\ref{Imax}) provide good estimates for both the positions and the heights of the first peak, as can be seen in Fig.~\ref{vacc_EI_evol}.

%%%%%%%%%%%%%%%%%%%%%%%%%%%%%%%%
\section{Cellular automaton simulations}\label {IV}

The compartmental $SEIRS$ model, considered in sections \ref{II}-\ref{III}, has a number of limitations. One of them is the assumption of perfect miscibility between the individuals compising the population. It is assumed, that during a discrete time step, any susceptible individual can meet any infective one and contract the disease with the constant probability $\beta$. Therefore, the latter is a composite probability, including: probability for two individuals to meet, probability that the infected individual speads virus around, and probability that the susceptable individual is infected. In real life, all three may be quite independent and subject to certain distributions. Namely, the individuals are distributed in space, leading to a certain geography-based arrangement, as well as constituting a network of social relations, resulting in a random, scale-free, of a small-world type structures \cite{Holovatch2017,Amati2018}. Furthemore, the probabilities that infected individual spreads virus, and that the susceptable one contracts it, both may be distributed by a certain law.

Here we consider one of the simplest cases, that takes into account spatial arrangement of individuals, namely: a simple square lattice with a neighbourhood size for each individual, given by the number $q$ of its neighbours. The setup is similar to that being discussed for the case of a simpler $SIS$ model in Ref.~\cite{Ilnytskyi2016}. In particular, the vertex set of a graph is $\mathbb{Z}^2$, and each individual is characterized by a neighbourhood radius $R_q>0$, such that the neighbourhood of a given $k \in \mathbb{Z}^2$ is defined as: $k' \sim k$ whenever $|k'-k|\leq R_q$. The choice of $R_q$ is made to obtain the neighbourhood size equal to chosen neighbourhood size $q$. Eeach $k$ is associated with an individual, that is characterized by its state $s_k$, which takes any value from a set of $\{s,e,i,r\}$, the letters correspond to the susceptible, unidentified infected, identified infected and recovered states, respectively. The evolution of the system of $N$ such individuals is driven by the algorithm, based on transition rules between the states of each individual in the $SEIRS$ model, depicted schematically in Fig.~\ref{Model}. The principal difference with the compatmental realization of this model is that the infecting occurs locally, within the sphere of the radius $R_q$ around each infected individual. 

We used the following parameters: $L=700$, and, consequently, the total number of individuals is $N=L^2=490\,000$, which brings the model to a realistic size for a medium-sized town. The periodic boundary conditions are applied along both Cartesian axes. The set of neighbourhood sizes covers cases from $q=N$ (a limit, that reproduces the compartmental model) down to $q=4$ (nearest neighbours only von Neumann neighbourhood), this allows to consider the effect of quarantine explicitly: by reducing the $q$ value. At each time instance, we made $N$ random choices of individuals with their current states updated according to their current state and the states of theit neighbours. This levels the time discretization with the $\Delta t$, used for the numeric solution (\ref{integr}) of the compartmental model. We employed the asynchronous realization of the cellular automaton algorithm, where the change of $s_k$ is immediate and affects the rest of $N$ updating attempts, performed at a given time instance.

We focus here on the time evolution of the fraction of identified infected individuals, $I$, because it provides the level of load on the medical system of an imaginary town. As remarked above, spatial arrangement of individuals on sites of a square lattice allows explicit modelling of the quarantine measures, via the neighbourhood size $q$. We choose the option, when this size is the same for all $N$ individuals, but other cases are equally possible. In particular, $q$ can be chosen randomly within a certain interval of values $[0:q_{\mathrm{max}}]$ \cite{Ilnytskyi2016}, or from a given distribution. In the latter case, one can mimic certain network structures in an implicit way.

\begin{figure}[!ht]
\begin{center}
\includegraphics[clip,width=12.3cm,angle=0]{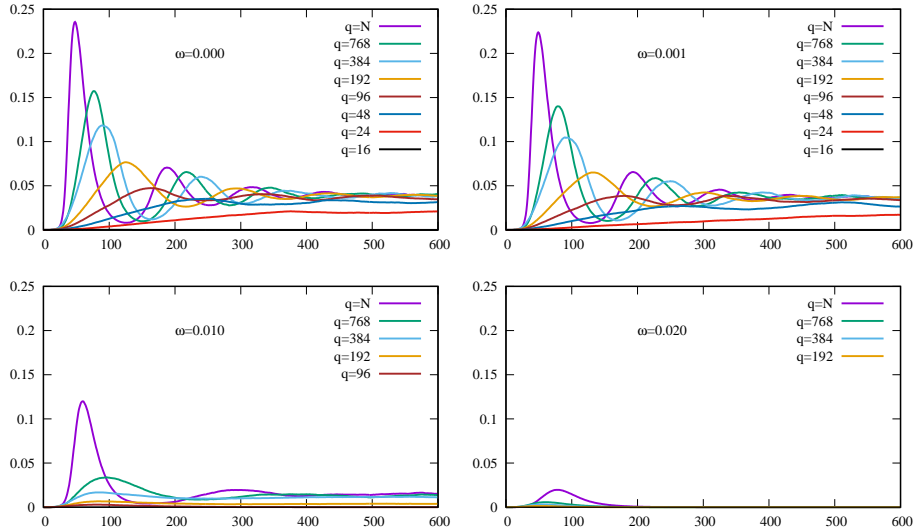}
\caption{\label{CA_I_evol}The effect of vaccination rate $\omega$ and of the quarantine measures (via the neighbourhood size $q$) on time evolution of the fraction $I(t)$. The case of $\beta=0.5$, $\alpha=0.1$ is shown for the system of $N=490\,000$ individuals on a square lattice with the system evolution driven via cellular automaton algorithm.}
\end{center}
\end{figure}
Fig.~\ref{CA_I_evol} shows the time evolution $I(t)$ for four cases: no ($\omega=0$), slow ($\omega=0.001$), fast ($\omega=0.010$) and ultrafast ($\omega=0.020$) vaccination at fixed $\beta=0.5$, $\alpha=0.1$. In each case we consider the effect of introducing the quarantine measures, via lowering the neighbourhood size $q$. For the no vaccination case, the quarantine measures resulted in two effects: the first is a shift of a first and consequent peaks to longer times, and the second is lowering their maximum values. Let us note, that the stationary state (at least, its estimate at large $t\sim 600$) is still a nonzero value $I(600)>0$, until very strict quarantine measures ($q=16$) are undertaken. Of course, such measures affect the society strongly, both economically and in a mental way. The situation is quite similar for the slow vaccination case ($\omega=0.001$), where small shifts of all the peaks to the right and small decrease of their maxima are observed. The behaviour of $I(t)$ changes drastically for the fast vaccination case ($\omega=0.010$), where one observes not only essential lowering of all peaks, but also a decrease of the $I(600)$ value. Supression of the epidemic outbreak can be achieved at much relaxed quarantine measures, the neighbourhood size of $q=768$ instead of $q=48$ for the no vaccination case. This tendency stenthens further when the vaccination rate is increased to $\omega=0.020$. In this case, even in the case of no quarantine measures implemented, the outbreak does not progress beyond $I\approx 0.02$ for the current choice of the $\beta$ and $\alpha$ rates. Similar results are obtained for the other choices of $\beta$ and $\alpha$, these are not shown for the sake of brevity.

\begin{figure}[!ht]
\begin{center}
\includegraphics[clip,width=8cm,angle=0]{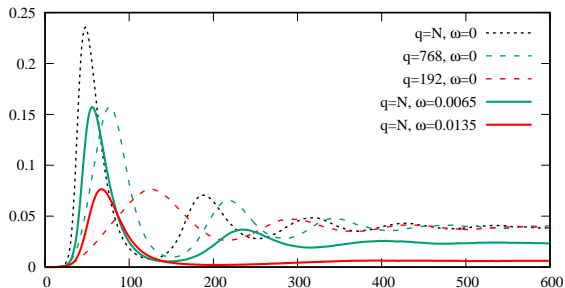}
\caption{\label{CA_I_match}An attempt to match the effect of vaccination (different $\omega$ rates) and that of the quarantine measures (different neighbourhood sizes $q$), as seen on time evolution of the fraction $I(t)$. The case of $\beta=0.5$, $\alpha=0.1$ is shown for the system of $N=490\,000$ individuals on a square lattice with the system evolution driven via cellular automaton algorithm.}
\end{center}
\end{figure}
In a real life, one would attempt to balance both measures, to achieve efficient suppression of an outbreak by means of minimal cost. To this end we tried to match the cases, when only one, either vaccination, or quarantine, measure is implemented, for the same case of $\beta=0.5$ and $\alpha=0.1$. The result is shown in Fig.~\ref{CA_I_match}, where we matched the height of a first peak in both cases. In this way, the vaccination rate $\omega=0.0065$ can be associated with the quarantine measures of $q=768$, whereas the vaccination rate $\omega=0.0135$ with $q=192$, respectively. However, one should remark, that equivalent vaccination rate, found in this way, does not shift the position of the first peak (observed by implementing equivalent quarantine measures), and leads to much stronger reduction of the height of the second peak as compared with equivalent quarantine measures. Therefore, the exact match between two measures is impossible, but, nevertheless, the results given in Figs.~\ref{CA_I_evol} and \ref{CA_I_match} provide some trends and tendencies that may help in balancing between these two measures.

%%%%%%%%%%%%%%%%%%%%%%%%%%%%%%%%
\section{Conclusions}\label {V}

We propose here the $SEIRS$ epidemiology model, which takes into consideration principal features of the COVID-19 pandemic: abundance of unidentified (asymptomatic or with mild symptoms) infected individuals, limited time of immunity and a possibility of vaccination. It assumes four states for individuals: susceptible, unidentified infected, identified infected and recovered with temporal immunity. Vaccination changes the state from susceptible to recovered. Model involves: the virus transfer $\beta$, identification $\alpha$, and vaccination $\omega$ rates, all can be varied, as well as fixed curing $\gamma$ and loss of immunity $\varphi$ rates based on an average statistics.

For the case of a compartmental realization of this model, we found two stationary states (fixed points) for the set of differential equations of the model: the disease-free and the endemic one. They exist in their respective restricted regions of the parameter space because of the limitations for all fractions to stay positive and do not exceed $1$. The linear stability analysis indicates the stability of both fixed points within their respective regions. The expression of the basic reproductive number enables to discuss the ways to bring the number of infected individuals in a stationary state down via changing the model parameters. This can be achieved by lowering the contact rate $\beta$ and/or by the increase of the identification $\alpha$ and vaccination $\omega$ rates. However, if $\beta$ equals or exceeds the critical value $\beta_c$, the disease-free fixed point can not be achieved at any combination of $\alpha$ and $\delta$. 

Numeric integration is employed to obtain the time evolution for the fractions of infected individuals. The results for the positions and heights of a first peak for the fractions of infected individuals, obtained numerically, are fitted to simple algebraic forms. These are based on observations, that the positions of the peaks diverge and their heights turn into zero, when the identification rate $\alpha$ approaches the critical value $\alpha_c$. Obtained algebraic expressions provides means for simple estimates for the first peak positions and height depending on the set of values for model rates.

To evaluate the effect of quarantine measures explicitly, we performed computer simulations of a lattice-based realization for this model using the cellular automaton algorithm. We found, that the quarantine measures delay and lower the first and subsequent peaks for the fraction of infected individulas, but do not lead to the complete elimination of the disease at long times, unless extremely restrictive measures are taken. On contrary, vaccination affects both the peak heights and the long time values for this fraction. The attempt is made to match both quarantine and vaccination measures aimed on balanced solution for effective suppression of the pandemic.       

%%%%%%%%%%%%%%%%%%%%%%%%%%%%%%%%
\section{Acknowledgements}

This work was supported by the National Research Foundation of Ukraine (Project No. 2020.01/0338).
The computer simulations have been performed on the computing cluster of the Institute for Condensed Matter Physics of NAS of Ukraine (Lviv, Ukraine).

%%%%%%%%%%%%%%%%%%%%%%%%%%%%%%%%
\bibliography{SEIR_covid.bib}{}
\end{document}